\documentclass[twocolumn]{aastex63}
\received{2020 June 23}
\revised{2020 July 23}
\accepted{2020 July 23}
\submitjournal{ApJ}

\shorttitle{On the formation of Lyman~$\beta$ and the O~{\sc i} 1027 and 1028 \AA}
\shortauthors{Hasegawa et al.}

\begin{document}

\title{On the formation of Lyman~$\beta$ and the O~{\sc i} 1027 and 1028 \AA\ spectral lines \footnote{Released on ---}}

\correspondingauthor{Takahiro Hasegawa}
\email{hasegawa.takahiro@ac.jaxa.jp}

\author[0000-0003-3306-9525]{Takahiro Hasegawa}
\affiliation{Department of Earth and Planetary Science, The University of Tokyo, 7-3-1, Hongo, Bunkyo-ku, Tokyo 113-0033, Japan}
\affiliation{Institute of Space and Astronautical Science, Japan Aerospace Exploration Agency, 3-1-1, Yoshinodai, Chuo, Sagamihara, Kanagawa 252-5210, Japan}

\author[0000-0001-5518-8782]{Carlos Quintero Noda}
\affiliation{Rosseland Centre for Solar Physics, University of Oslo, P.O. Box 1029 Blindern, N-0315 Oslo, Norway}
\affiliation{Institute of Theoretical Astrophysics, University of Oslo, P.O. Box 1029 Blindern, N-0315 Oslo, Norway}
\affiliation{Instituto de Astrof\'isica de Canarias, E-38200, La Laguna, Tenerife, Spain.}
\affiliation{Departamento de Astrof\'isica, Univ. de La Laguna, La Laguna, Tenerife, E-38205, Spain}

\author[0000-0003-4764-6856]{Toshifumi Shimizu}
\affiliation{Institute of Space and Astronautical Science, Japan Aerospace Exploration Agency, 3-1-1, Yoshinodai, Chuo, Sagamihara, Kanagawa 252-5210, Japan}
\affiliation{Department of Earth and Planetary Science, The University of Tokyo, 7-3-1, Hongo, Bunkyo-ku, Tokyo 113-0033, Japan}

\author[0000-0001-9218-3139]{Mats Carlsson}
\affiliation{Rosseland Centre for Solar Physics, University of Oslo, P.O. Box 1029 Blindern, N-0315 Oslo, Norway}
\affiliation{Institute of Theoretical Astrophysics, University of Oslo, P.O. Box 1029 Blindern, N-0315 Oslo, Norway}



\begin{abstract}

We study the capabilities of Lyman~$\beta$ and the O~{\sc i} 1027 and 1028~\AA\ spectral lines for understanding the properties of the chromosphere and transition region. The oxygen transitions are located in the wing of Lyman~$\beta$ that is a candidate spectral line for the solar missions Solar Orbiter/SPICE and Solar-C (EUVST). We examine general spectroscopic properties of the three transitions in the quiet Sun by synthesizing them assuming non-local thermal equilibrium taking into account partial redistribution effects. We estimate the heights where the spectral lines are sensitive to the physical parameters computing the response functions to temperature and velocity using a 1D semi-empirical atmospheric model.  We also synthesize the intensity spectrum using the 3D enhanced network simulation computed with the {\sc Bifrost} code. The results indicate that Lyman~$\beta$ is sensitive to the temperature from the middle chromosphere to the transition region while it is mainly sensitive to the line-of-sight velocity at the latter atmospheric layers, around 2000~km above the optical surface. The O~{\sc i} lines form lower in the middle chromosphere, being sensitive to the LOS velocities at lower heights than those covered by Lyman~$\beta$. The spatial distribution of intensity signals computed with the {\sc Bifrost} atmosphere, as well as the inferred velocities from the line core Doppler shift confirm the previous results. Therefore, these results indicate that the spectral window at 1025~\AA\ contains several spectral lines that complement each other to seamlessly trace
the thermal structure and gas dynamics
from the middle chromosphere to the lower transition region. 
\end{abstract}



\section{Introduction} \label{sec:intro}

The chromosphere is an inhomogeneous and very dynamic region located in between the solar photosphere and the corona. It is the place where radiative equilibrium breaks down with a temperature that is higher than that at the upper photosphere, and it is the key element for better understanding most of the hot topics in solar and stellar atmospheres. 
For instance, it is in the chromosphere where we have the transition from local thermodynamic equilibrium (LTE) to non-LTE for the radiative transfer, or from plasma- (photosphere) to magnetic field-dominant (corona) regimes. 
Furthermore, from temperature minimum to the corona, the state of plasma changes from neutral to partially- and fully-ionized.
Therefore, a better knowledge of the chromosphere can provide an essential understanding of universal astrophysical processes as magnetic reconnection or multi-fluid effects \citep[see, for instance, the recent review of][]{Carlsson2019}.

Several candidate spectral lines form in the chromosphere and the transition region (TR), for instance, the H transitions from the Lyman series. The resonance line Lyman~$\alpha$ at 1216\AA \ has the strongest intensity among emission lines in the extreme ultra-violet (EUV) regime, followed by Lyman~$\beta$ at 1025~\AA. These transitions were investigated in the past through observations taken by the Laboratoire de Physique Stellaire et Plan\'{e}taire (LPSP) instrument on board the OSO-8 satellite \citep{Lemaire1978}. Later on, regular observations of Lyman spectral lines were done with the SUMER~instrument \citep{Wilhelm1995, Lemaire1997} onboard the SoHO spacecraft~\citep{Domingo1995}. Among the vast number of works the mission produced, we have that \cite{Zhang2010} studied the behavior of Lyman~$\beta$ in TR explosive events, with lines of O~{\sc vi} at 1032 and C~{\sc ii} at 1037~\AA. \cite{Xia2004} studied some spectral lines ranging from chromospheric to low-coronal temperature, including the transitions mentioned above, to investigate the relationship between, e.g., the spectral lines Doppler shift and photospheric magnetic field in network regions. \cite{Tian2009} compared the Lyman~$\alpha$ and Lyman~$\beta$ lines in the chromospheric network and the inter-network. Some authors, e.g., \cite{Heinzel2001} and \cite{Gunar2007}, also used the Lyman transitions to study prominences. 

In the case of the Lyman~$\alpha$ observations with SUMER, the line intensity was so strong that an attenuator was used to avoid the saturation of the detector  \citep{Wilhelm1995}. However, this led to some unexpected alteration of the line profiles and additional difficulties to analyze the Lyman~$\alpha$ observations. On the other hand, Lyman~$\beta$ did not have such a problem and was observed more frequently. We assume this could be a reason why Lyman~$\beta$ is one of the selected spectral lines of the Spectral Imaging of the Coronal Environment (SPICE) instrument \citep{SPICE2019} onboard Solar Orbiter (SO, \citealt{Muller2013}). Furthermore, the future Solar-C (EUVST) mission \citep{Shimizu2019} will also have access to Lyman~$\beta$ (Lyman~$\alpha$ too in this case) observing the solar spectrum at EUV wavelengths. In particular, the latter mission aims to have access to different atmospheric layers from the chromosphere to the outer corona. 

Coronal lines that fall in the EUV part of the spectrum are well known from past missions, like SoHO/SUMER~\citep{Domingo1995,Wilhelm1995, Lemaire1997} or Hinode/EIS~\citep{Kosugi2007, Culhane2007}. However, the same cannot be said for chromospheric and TR lines in the mentioned spectral range. There are some prominent exceptions, such as the Mg~{\sc ii} $h$ \& $k$ lines that are observed by the Interface Region Imaging Spectrograph (IRIS, \citealt{DePontieu2014}). Unfortunately, although Solar-C (EUVST) is designed to obtain Mg~{\sc ii} filtergrams, the Mg~{\sc ii} transitions are out of the scope of the EUV spectroscopic missions mentioned before. 

In this work, we aim to start a series of publications similar to the works entitled "The formation of IRIS diagnostics" \citep[][]{Leenaarts2013a,Leenaarts2013b,Pereira2013,Pereira2015,Rathore2015,Rathore2015a,Lin2015,Rathore2015b,Lin2017} to support future missions. We believe that performing similar studies for SPICE, Solar-C (EUVST), and future EUV instruments will help the missions to perform multi-wavelength observations of the middle and upper atmosphere optimally.  In this publication, we examine the properties of the hydrogen Lyman~$\beta$ spectral line and the doublet of O~{\sc i} 1027 and 1028~\AA \ lines located in the wing of Lyman~$\beta$. The properties of the O~{\sc i} transitions are not known in detail, despite being located in the red wing of Lyman~$\beta$ and they have been used mainly as a reference for wavelength calibrations in SoHO/SUMER observations. However, we believe they are good candidates to infer the atmospheric properties at lower heights than those covered by Lyman~$\beta$, complementing its capabilities. Most importantly, they are located at less than 2~\AA \ from the Lyman~$\beta$ core, so it is feasible to scan the three of them simultaneously. In the following, we perform various theoretical studies aiming at determining the possibilities of the Lyman~$\beta$ and the O~{\sc i} lines for comprehending the solar atmosphere.

This paper is organized as follows. In Section 2, we show the methodology and models we use in this work. Section 3 describes the results of some tests and of computation with a realistic 3D atmosphere. Section 4 shows discussions, and we summarize this work in Section 5.


\section{Methodology} \label{sec:method}

\subsection{Spectral lines}

\begin{figure}
\centering
\includegraphics[trim=0 0 0 0,width=8.5cm]{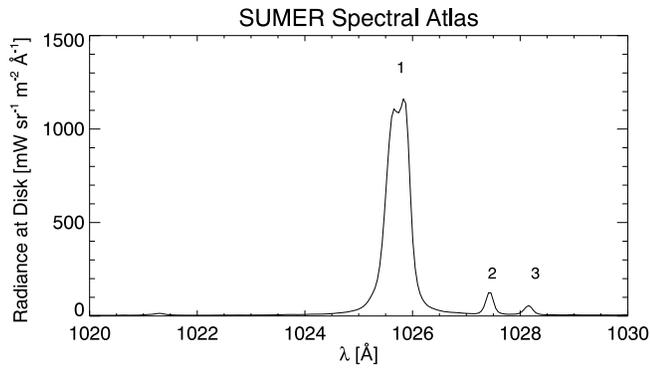}
\caption{Solar EUV spectrum at 1025~\AA\ observed by the SUMER instrument onboard SoHO. The observation corresponds to the SUMER quiet Sun atlas presented in \citet{Curdt2001}. We label the lines with numbers described in Table~\ref{table:lines}. \label{fig:atlas}}
\end{figure}

Figure~\ref{fig:atlas} depicts the quiet Sun atlas in the neighbourhood of Lyman~$\beta$. It was observed by SoHO/SUMER and presented in \cite{Curdt2001}. The red wing of the Lyman~$\beta$ encompasses two bound-bound transitions from neutral oxygen \citep{Wiese1996a} located at 1027 and 1028~\AA. In the case of the O~{\sc i} 1027~\AA, we have 2 spectral lines separated by 0.74 m\AA\ that are generated from the transition between 2 levels of the term ${\rm 2p^3~3d~{}^3D^o}$. In particular, the level with $J=1$ and energy 97488.378~$\rm cm^{-1}$ and the level with $J=2$ and energy 97488.448~$\rm cm^{-1}$, with the lower level ${\rm  2p^4~{}^3P_1}$ and energy 158.265~${\rm cm^{-1}}$ (see labels $2a$ and $2b$ in Table~\ref{table:lines}). In general, the mentioned wavelength separation is much smaller than the spectral resolution of EUV instruments onboard solar missions, as Hinode/EIS, SoHO/SUMER, IRIS, SO/SPICE or Solar-C (EUVST). Therefore, in practice, those transitions are observed as a single line which, from now on, we name the 1027 line.

The O~{\sc i}  1028~\AA\ spectral line corresponds to the transition between the level of the term ${\rm 2p^3~3d~{}^3D^o}$ with $J=1$ and energy 97488.378~$\rm cm^{-1}$, and the lower level ${\rm 2p^4~{}^3P_0}$ and energy ${\rm 226.977~cm^{-1}}$ (see label 3 in Table~\ref{table:lines}). Hereafter, we refer to this line as the 1028 line.

\begin{table}
\centering
\caption{Spectral information of the spectral lines}
\begin{tabular}{cccccccc} \hline
	Label  & Atom  & $\lambda$ [\AA]  & E$_{\rm l}$ ${\rm [cm^{-1}]}$ &  E$_{\rm u}$ ${\rm [cm^{-1}]}$ 	\\  \hline 
	1 & H~{\sc i} &1025.72 & 000.000 & 97492.28\\
	2a & O~{\sc i}& 1027.43  & 158.265 & 97488.378\\
	2b & O~{\sc i}& 1027.43 &  158.265 & 97488.448\\
	3 & O~{\sc i}& 1028.15 &  226.977 & 97488.378\\
\hline
\end{tabular}
\tablecomments{From left to right: the labels used in this work, the corresponding atomic species, the rest wavelength, and the energy of the lower (${\rm E_ l}$) and upper (${\rm E_u}$) levels of the transitions, respectively.}
\label{table:lines}
\end{table}

\subsection{Model atom} \label{subsec:atom}

We employ the O~{\sc i} 16 level atom model presented in \cite{Lin2015} for the computation of the oxygen lines of interest. In addition, the Lyman~$\beta$ emission populates the O~{\sc i} ${\rm 2p^3~3d~{}^3D^o}$ term through the O~{\sc i} 1025.76\AA\ spectral lines (corresponding to the transition between O~{\sc i} ${\rm 2p^3~3d~{}^3D^o}$ and the O~{\sc i} ground state). This atomic state corresponds to the upper term of the 1027 and 1028 \AA\ spectral lines. Due to the over-population, O~{\sc i} line intensities are enhanced by cascade (Lyman~$\beta$ pumping, \cite{Bowen1947}). In the works of \cite{Carlsson1993} and \cite{Lin2015}, the authors explain that the Lyman~$\beta$ pumping is a major source of excitation of the O~{\sc i} resonance lines, and therefore, in order to properly compute those transitions we should also compute the atomic populations of H~{\sc i}. We use a simplified hydrogen atom, found inside the RH libraries, consisting of 6 levels, 10 bound-bound transitions, and 5 bound-free transitions, including the Lyman series (from $\alpha$ to $\delta$) transitions. 

In the wavelength range of interest (around 1025\AA), the carbon opacity actively contributes to the continuum, and it is subject to NLTE scattering. We confirmed that computing the population of C in NLTE reduces the continuum intensity. However, since we focus on the properties of emission lines, we treat C in LTE in this work.

\begin{figure*}
\centering
\includegraphics[trim=0 0 0 0,width=17.5cm]{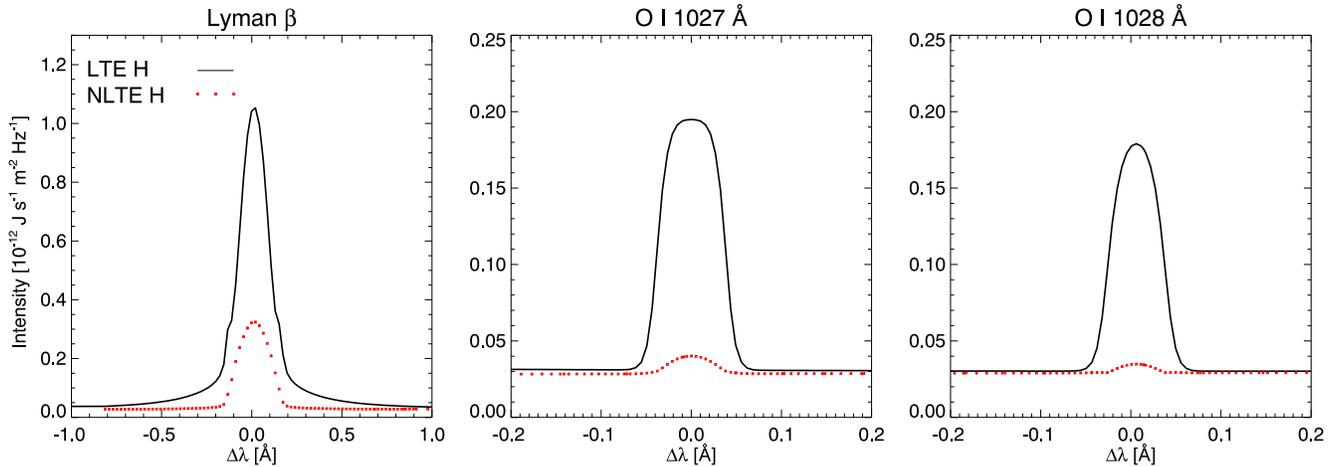}
\caption{Intensity profiles for the Lyman~$\beta$, 1027, and 1028 spectral lines computing the hydrogen populations in LTE (black) or in NLTE (red). In the three cases, the oxygen is treated in NLTE.  \label{fig:test2}}
\end{figure*}

\subsection{Model Atmosphere}

We use two types of atmospheric models. We start with the semi-empirical FALC atmosphere \citep{Fontenla1993} to determine the optimum configuration for synthesizing the spectral lines of interest. We also use it to examine sensitivity of the spectral lines to perturbations on the atmospheric parameters. Later, we employ the snapshot 385 of the {\sc Bifrost} \citep{Gudiksen2011} enhanced network simulation \citep{Carlsson2016}. This atmosphere covers from the upper convection zone to the lower corona with a total size of $24\times24\times16.8$~Mm composed of $504\times504\times496$ grid cells equidistant in the horizontal direction (48~km cell size) and non-uniformly spaced in the vertical direction. The simulation has been used in several works related to the synthesis of spectral lines \citep[e.g.,][]{Leenaarts2015,Stepan2016, QuinteroNoda2016,Sukhorukov2017}. To reduce the computational time we cut the vertical domain that in this work goes from z=[-0.5,5]~Mm, being z=0~Mm the height where $\tau=1$ in average for the continuum at wavelength 5000~\AA. Besides that change, we use the original snapshot for this study, i.e. we do not apply any spatial degradation nor instrumental effects on the synthetic spectra. 

\subsection{Synthesis of the emergent intensity} \label{synthesis}

We solve the radiative transfer and statistical equilibrium equations using the RH code \citep{Uitenbroek2001}. The code allows us to compute the atomic populations of the different levels associated with given transitions considering non-local thermodynamic equilibrium (NLTE). Moreover, we can assume complete redistribution (CRD) or take into account partial redistribution (PRD) effects for any transition. We start this work computing the intensity profiles for different scenarios under both regimes, i.e., PRD and CRD, using the FALC atmosphere. After that we study the spatial distribution of intensity signals computed with the {\sc Bifrost} atmosphere. For both type of atmospheric models, we assume that the atmosphere is plane-parallel using the 1D geometry package of RH. We believe that, for this first study, it is reasonable to work under that approximation.  However, we plan to examine the impact of 3D effects for these lines, in particular for Lyman~$\beta$, in the future using, e.g., the 3D geometry package of RH, or additional codes like Multi3D \citep{Leenaarts2009Multi3D} or PORTA \citep{Stepan2013}.


\section{Results} \label{sec:result}

\subsection{Simulation Setup} \label{sec:setup}

Before discussing the diagnostic potential of the spectral lines, we need to determine the optimum configuration for synthesizing the spectrum, e.g., whether we consider non-LTE processes for all lines and whether PRD needs to be included as well. Our target is to define a computationally efficient, yet sufficiently accurate, synthesis configuration for use in the following computations, in particular, for the 3D {\sc Bifrost} simulations.

\subsubsection{NLTE effects}

We first examined the validity of the LTE approximation for the O~{\sc i} spectral lines by comparing the differences between the synthetic profiles in LTE and NLTE. We found for both transitions a deviation of up to 10 times, indicating that we need to synthesize the spectral lines in NLTE. The next step is to study the impact on the 1027 and 1028 transitions of solving or not the atomic populations of neutral hydrogen. As mentioned before, \cite{Carlsson1993} and \cite{Lin2015} showed that the pumping process of Lyman~$\beta$ could modify the population state of the energy levels of O~{\sc i}. Moreover, the 1027 and 1028 spectral lines are located in the wing of Lyman~$\beta$.

For this study, we always assume NLTE when computing the populations of oxygen, and switch between LTE and NLTE for the hydrogen populations. We consider CRD for now, and we compute the emergent intensity for $\mu = 1$ (where $\mu = \cos{\theta}$ and $\theta$ is the heliocentric angle, i.e., the angle between the ray and the normal of the atmosphere).

Figure~\ref{fig:test2} shows the profiles of Lyman~$\beta$, 1027, and 1028 lines in the cases where we treat hydrogen in LTE (black) and NLTE (red). 
By considering NLTE, the line core intensity of Lyman~$\beta$ is reduced to around 1/3 of that in LTE. We also see a reduction of the line core intensity for the 1027 and 1028 lines, up to about 1/5 of that obtained in LTE. Therefore, we need to solve both, neutral hydrogen and oxygen populations in NLTE to properly synthesize the 1027 and 1028 spectral lines.

\subsubsection{PRD Effects of Lyman lines} \label{sec:hac}

\begin{figure*}
\centering
\includegraphics[trim=0 0 0 0,width=17.5cm]{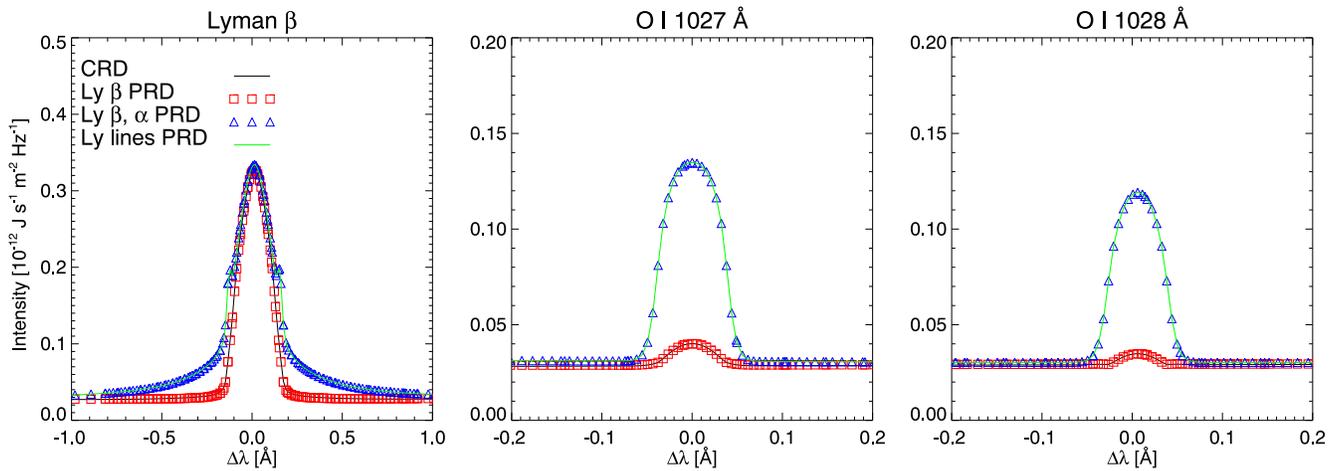}
\caption{PRD effects on the synthetic, from left to right, Lyman~$\beta$, 1027, and 1028 lines. We study four different configurations for the hydrogen transitions; 1) all Lyman transitions are treated in CRD (solid black), 2) only Lyman~$\beta$ is computed in PRD (red squares), 3) Lyman~$\beta$ and $\alpha$ are treated in PRD (blue triangles), and 4) all Lyman transitions included in the atom model (Lyman $\alpha, \beta, \gamma$, and $\delta$) are computed in PRD (solid green). All the configurations correspond to a heliocentric angle of $\mu=1$. \label{test3}}
\end{figure*}

Based on the results presented in the previous sections, we want to examine the impact on the 1027 and 1028 lines of considering PRD for the hydrogen transitions. We start from the previous configuration, i.e., H and O populations are computed in NLTE, and we compare four different cases. First, all Lyman transitions are synthesized in CRD. Second, only Lyman~$\beta$ is computed taking into account PRD effects while the rest of the transitions are treated in CRD. Third, we compute Lyman~$\beta$ and $\alpha$ assuming PRD and the rest of transitions in CRD. Fourth, all Lyman series transitions (i.e., Ly$\alpha, \beta, \gamma$, and $\delta$) are computed considering PRD effects. In addition, for now, oxygen transitions are all treated in CRD.

Figure~\ref{test3} shows the results of these studies where the four mentioned cases can be divided into two groups; when Lyman~$\alpha$ is computed in CRD (the 1st and 2nd cases) or not (the 3rd and 4th cases).  Including PRD in the computation of Lyman~$\beta$ slightly modifies the spectral properties (red squares). However, when Lyman~$\alpha$ is treated under PRD, the differences are more noticeable for the three spectral lines (blue triangles). Those differences do not increase when including PRD for the rest of Lyman transitions (solid green). Thus, we can consider PRD effects only on Lyman~$\alpha$ and $\beta$ for computing the spectral lines of interest while the rest of the transitions can be treated assuming CRD. Doing this, we also manage to save computational time, the latter configuration being around 10\% faster than the case where all the Lyman transitions are treated in PRD.

\subsubsection{PRD effects of the O~{\sc i} lines} \label{sec:hps}

We examine the impact of PRD on the O~{\sc i} 1027 and 1028~\AA \ transitions in this section. We consider PRD for all the transitions included in our O~{\sc i} model atom and compare the results with the CRD case. We use the FALC atmosphere again, and we compute the synthetic profiles for different heliocentric angles. The differences between both computations are small being up to 4\% and 1\% for the 1027 and 1028 lines, respectively, in the extreme case of $\mu$=0.05. Thus, we opt to compute the O~{\sc i} transitions from now on in CRD, which will reduce the computational effort. In this case the computational savings are more noticeable than in the case of H if we take into account that the PRD computation is, in general, around 1.5 times slower than the case where all transitions are assumed in CRD.

\subsection{Line formation in a semi-empirical 1D atmosphere} \label{subsec:1d}

The first part of this publication was focused on finding an appropriate configuration for synthesizing the O~{\sc i} 1027 and 1028~\AA\ transitions. From now on, we study the properties of these lines using the semi-empirical FALC atmosphere.

\subsubsection{Height of formation} \label{subsec:tau}

\begin{figure}
\centering
\includegraphics[trim=0 0 0 0,width=8.3cm]{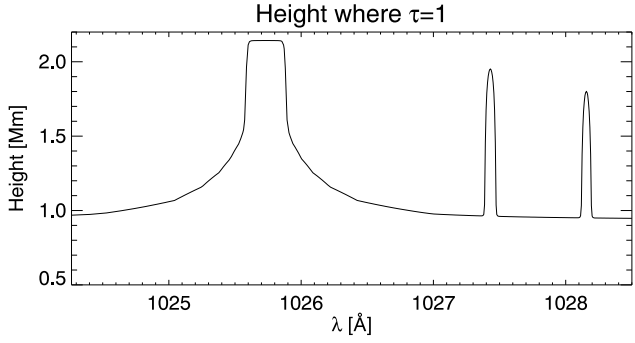}
\caption{Formation height of ${\rm Lyman \beta}$, O~{\sc i} 1027 and 1028~\AA \ given as the height where the optical depth is unity for the FALC atmosphere. \label{fig:height}}
\end{figure}

We estimate the formation height of the 1027 and 1028 lines computing where the optical depth is unity for the wavelength range presented in Figure~\ref{fig:atlas}. We show the results in Figure \ref{fig:height}. The EUV continuum forms around 1000~km while the Lyman~$\beta$ line goes up 2150~km that corresponds to the top boundary of the FALC atmosphere, i.e., transition region.
In the case of the O~{\sc i} spectral lines, they form at lower heights, in the upper chromosphere. Their height of formation is different, with 1027 reaching higher, around 1951~km, while 1028 forms approximately 200~km lower. Finally, it is worth mentioning that the height where the optical depth is unity at the line core wavelengths of Lyman~$\beta$ is flat for several m\AA. This behavior shows that the Lyman~$\beta$ transition reaches slightly higher than the top boundary of the FALC atmosphere.

\subsubsection{Response Function}\label{subsec:rf}

\begin{figure}
\centering
\includegraphics[trim=0 0 0 0,width=8.5cm]{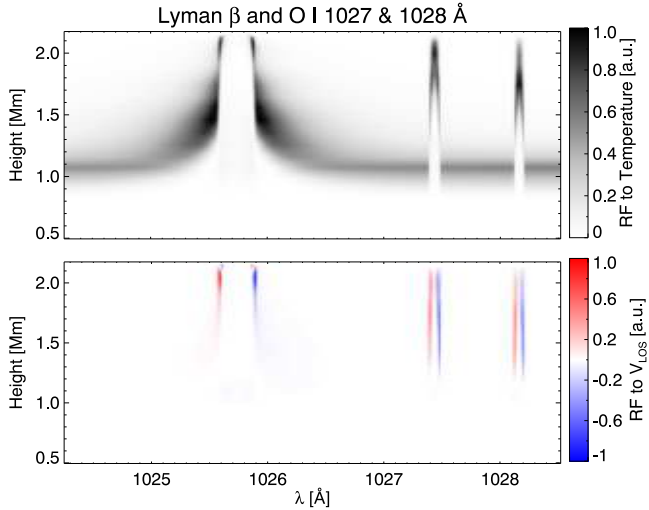}
\caption{Response functions to changes in the temperature (top) and the LOS velocity (bottom). Black or blue/red color corresponds to regions where the spectral lines are sensitive to a given perturbation at a certain height while white indicates no sensitivity to perturbations in the atmospheric parameters.  \label{fig:rf}}
\end{figure}

We continue using the FALC atmosphere, this time for estimating the sensitivity of the spectral lines of interest to perturbations on the atmospheric parameters. We compute the response functions (RF) ${\bf R(\lambda, z)}$ defined as
\begin{eqnarray}
\delta {\bf I}(\lambda)=\int_0^\infty {\bf R}(\lambda, \tau)\delta x\  dz.
\label{rf1}
\end{eqnarray}
where $\delta x$ is a perturbation we introduce in the atmosphere, $x$ is the atmospheric parameter that is modified, and $\delta {\bf I}(\lambda)$ is the impact those changes have on the synthetic profile \citep{Landi1977}.

We study the response of the spectral lines of interest to changes in temperature and line-of-sight (LOS) velocity. We compute the RF numerically after applying a perturbation constant with height and equal to 1 K and 100 ${\rm m\ s^{-1}}$ on the temperature and LOS velocity, respectively. 

Figure~\ref{fig:rf} displays the results for the Lyman~$\beta$, 1027 and 1028 spectral lines. The RF to temperature changes shows that the wings of Lyman~$\beta$ are sensitive to lower layers while the line core intensity is modified when we perturbed the upper part of the atmosphere. We do not see any indication of degeneracy, i.e. each wavelength point is only sensitive to a narrow range of atmospheric heights. The properties of the 1027 and 1028 lines are similar to Lyman~$\beta$ with no degeneracy and being most sensitive to perturbations at lower heights. Similar to the results of the previous section, those heights are located at 1995 and 1735~km.

In the case of the LOS velocity, the pattern is similar at higher layers, with Lyman~$\beta$ always forming higher than the other two spectral lines. Interestingly, we can see that the sensitivity of the wing of Lyman~$\beta$ at deeper layers is much weaker than that shown by the line core. This behavior is different from the RF to the temperature. We believe this effect is similar to what can be found for broad spectral lines like H$\alpha$ or the Ca~{\sc ii}~H \& K lines \citep[e.g.,][]{Chae2013}. However, we have that the LOS velocity RF for the 1027 and 1028 lines reach lower in the atmosphere indicating that they can complement the lack of sensitivity of Lyman~$\beta$ to the plasma velocity at those layers.

\subsection{Line formation in a realistic 3D atmosphere}

\begin{figure*}
\centering
\includegraphics[trim=0 -10 0 0,width=18cm]{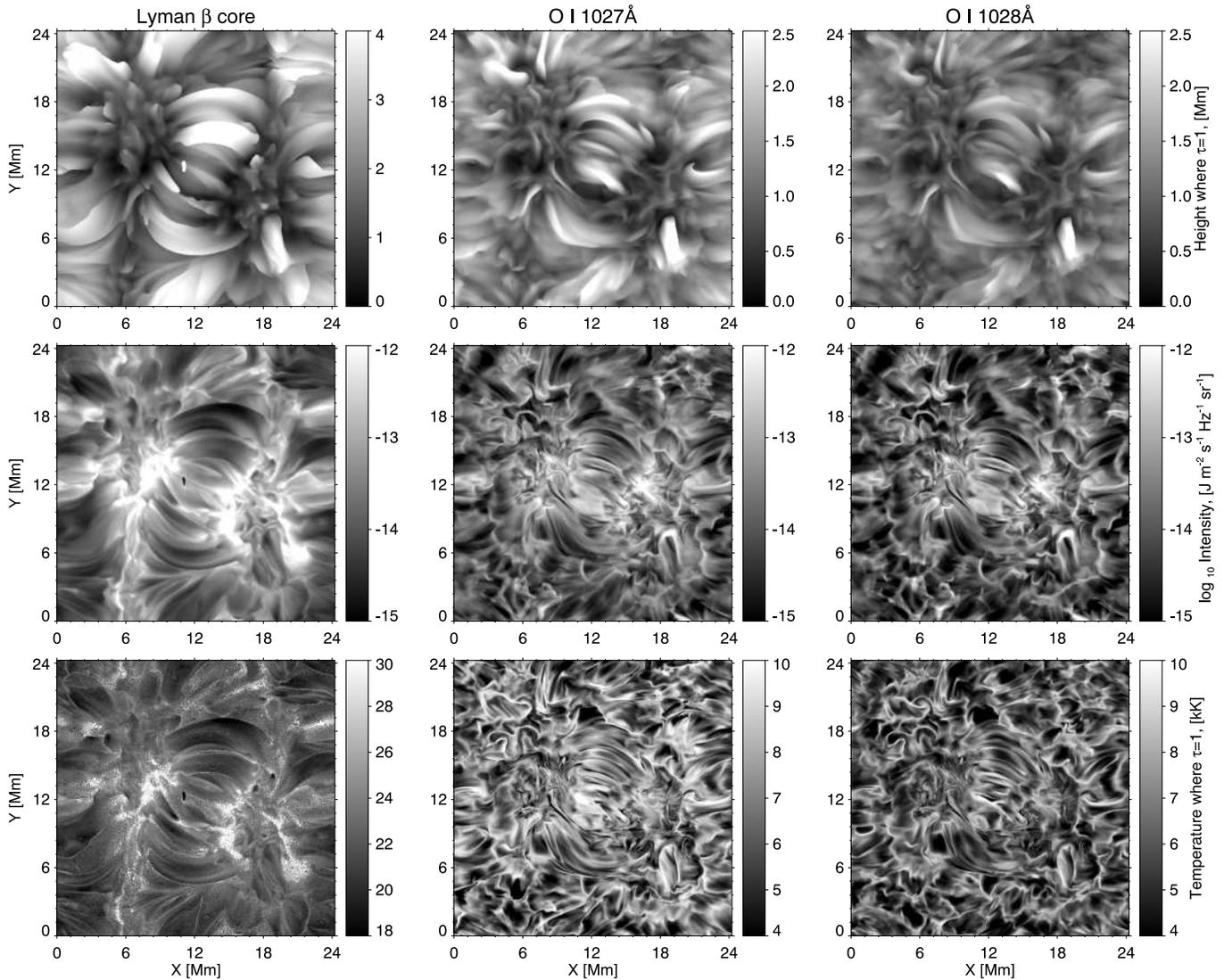}
\caption{First row shows the geometrical height where the optical depth is unity for the line core wavelengths of, from left to right, Lyman~$\beta$, 1027, and 1028 lines. Second row displays the intensity at the same wavelength location while the third row depicts the temperature at the heights presented in the upper row. \label{fig:bif_oi}}
\end{figure*}

\begin{figure*}
\centering
\includegraphics[trim=0 -10 0 0,width=18cm]{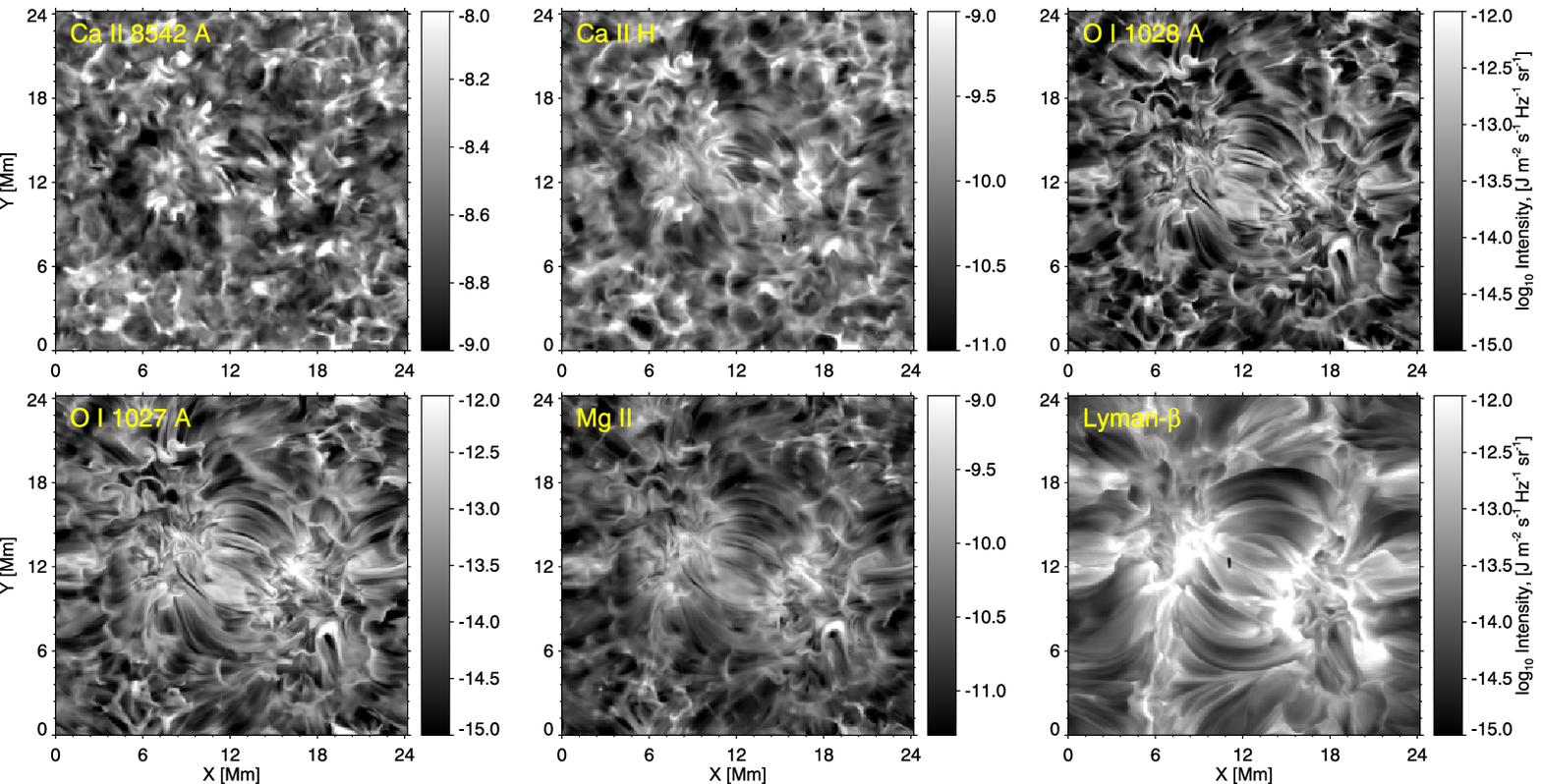}
\caption{ Spatial distribution of the line core intensity signals for a selection of chromospheric and transition region spectral lines. First row displays, from left to right, the results for Ca~{\sc ii} 8542~\AA \, Ca~{\sc ii} H, and O~{\sc i} 1028~\AA. Bottom row, from the left to right, shows  O~{\sc i} 1027~\AA, Mg~{\sc ii}~$k$ and ${\rm Ly\beta}$.\label{fig:bifrost}}
\end{figure*}

We study in this section the properties of the spectral lines for different spatial locations and atmospheric conditions using the atmosphere presented in \cite{Carlsson2016} and developed with the {\sc Bifrost} numerical code \citep{Gudiksen2011}. The simulation reproduces an enhanced network scenario where two magnetic field concentrations with opposite polarities dominate the field of view. We employ snapshot 385, the same that has been extensively used for various studies (e.g., \citealt{Leenaarts2013b,  Lin2015, Rathore2015, QuinteroNoda2016}).

We synthesize the spectrum performing column-by-column computations, i.e., we assume that the atmosphere is plane-parallel. We leave out from this work the possibility of performing 3D synthesis, something that was done in the past for H$\alpha$ \citep{Leenaarts2012} and Ca~{\sc ii} 8542~\AA \ \citep{Leenaarts2009, Stepan2016}. We also synthesize some of the most traditional chromospheric lines, i.e., Mg~{\sc ii} $h\&k$, Ca~{\sc ii} H\&K, and Ca~{\sc ii} 8542~\AA. They are among the commonly used candidates for analyzing the chromosphere and transition region, and they are observed in modern facilities like IRIS \citep{DePontieu2014}, Hinode \citep{Kosugi2007}, Sunrise \citep{Barthol2011}, CLASP-2 \citep{Narukage2016}, and ground-based telescopes like the Swedish 1-m Solar Telescope \citep{Scharmer2003, Scharmer2008}, the 4-m class telescopes Daniel K. Inouye Solar Telescope \citep{Keil2011} and the future European Solar Telescope \citep{Collados2013}. We computed the Mg~{\sc ii} $h\&k$  transitions assuming PRD with a simplified atom of 4 levels that contains only the transitions of interest. We compared the synthetic profiles with those generated by the atomic model presented in \cite{Leenaarts2013b}. The profiles from the simplified atom are slightly different, but we believe they are accurate enough for this comparison. The Ca~{\sc ii} spectral lines are synthesized from a 6 level atom similar to the one described in \cite{Shine1974} with Ca~{\sc ii} H\&K in PRD and the infrared triplet lines in CRD.

\subsubsection{Spatial distribution of intensity signals}

We show in Figure~\ref{fig:bif_oi} the results of the synthesis for the entire simulated field of view. The first row displays the spatial distribution of the geometrical height where the optical depth is unity for the line core wavelength of Lyman~$\beta$ (left), 1027 (middle) and 1028 (rightmost column). Starting with the H transition, we see loop-like structures that reach higher layers (up to 4~Mm) in the center of the FOV (see also Fig. 11 of \cite{Carlsson2016}). These structures connect the two bipolar magnetic concentrations at around [X,Y]=[5,14] and [16,10]~Mm that characterize the enhanced network simulation. The oxygen lines show a similar spatial distribution, but the loop-like structures occupy smaller areas indicating that the lines form lower in the atmosphere. 

The line core intensity for the three spectral lines is plotted in the second row of Figure~\ref{fig:bif_oi}. We found that the shape of Lyman~$\beta$ at the line core wavelengths could be particularly complex in some pixels, showing multiple lobes like the Mg~{\sc ii} lines, what could make it difficult to determine the exact wavelength position of the line core. To suppress the uncertainties, we used the results of the previous study, plotting the intensity where the optical depth unity is highest. We followed the same method for the oxygen lines. If we start with the Lyman~$\beta$ transition, we can see that the same structures found in the height of formation appear in the intensity map. The regions where the line forms higher show lower intensity values while the deeper parts, co-spatial with the magnetic field concentrations, display higher intensity values. A similar behavior is found for the oxygen lines, presenting a good correlation with the height of formation map. Again, the spatial distribution of intensity signals seems to correspond to lower atmospheric layers with, for instance, smaller and narrower loops at the center of the FOV.

We aim to investigate which kind of thermal structures the spectral lines are sensitive to. We add in the third row of Figure~\ref{fig:bif_oi} the temperature at the height where the optical depth is unity for the line core wavelengths. We believe this is a representative and simple visualization, although we know the spectral lines do not form at a unique and single height. It is a simple exercise that we can perform because we have both the simulation and the synthetic profiles. We have that the spatial distribution of the temperature maps is similar to that of the line core intensity (middle row). The average temperature at $\tau=1$ for Lyman~$\beta$ is $\sim 23000$~K and that of 1027 and 1028 is $\sim 6850$ and $\sim 6200$~K, respectively.

\begin{figure*}
\centering
\includegraphics[trim=0 0 0 +0,width=18cm]{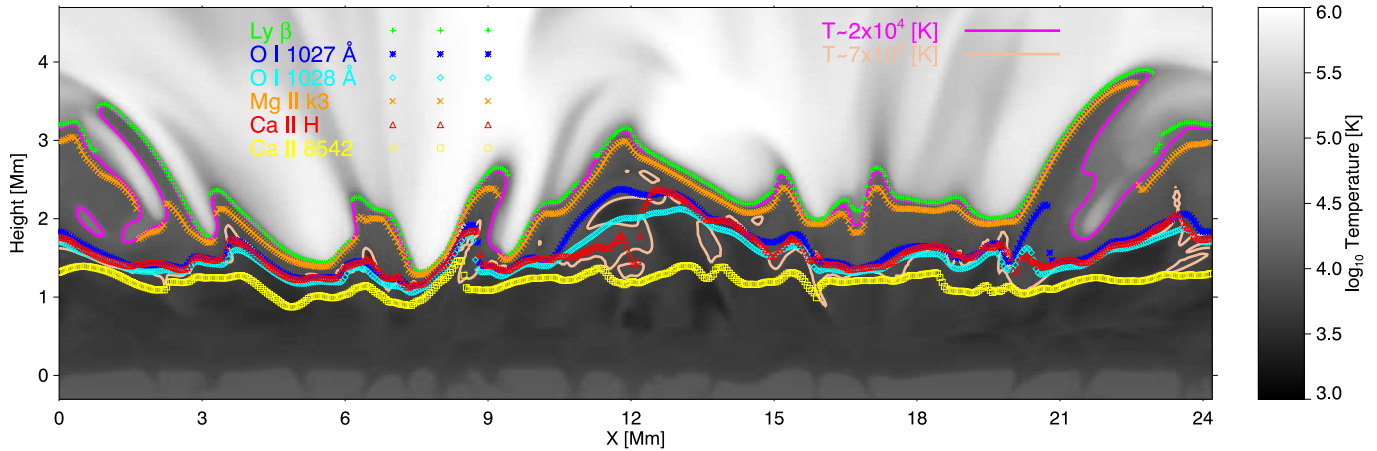}
\caption{Height where the optical depth is unity for the line core wavelengths of Lyman~$\beta$ (green), O~{\sc i} 1027(blue) and 1028~\AA \ (cyan). As a reference, we display the same height for Mg~{\sc ii}~$k$ (orange), Ca~{\sc ii} H (red), and Ca~{\sc ii} 8542~\AA \ (yellow). The background represents the gas temperature with isothermal contours at $2 \times 10^4$ (magenta) and $7 \times 10^3$ K (beige) and the cut corresponds to $Y=12$~Mm in Fig~\ref{fig:bif_oi}. \label{fig:heightbif}}
\end{figure*}

We present in Figure~\ref{fig:bifrost} the line core intensity for a selection of spectral lines that covers from the low-mid chromosphere (Ca~{\sc ii} 8542~\AA) to the transition region (Lyman~$\beta$). To compute those values, we take the intensity at the rest wavelength of each spectral line core. This approach is not accurate because in the presence of large velocity gradients, we are taking the wrong intensity value, but we believe it is accurate enough for most of the pixels and for this quick comparison. Starting with the infrared transition, we can see a faint indication of the connection between the opposite polarity magnetic footpoints (see the central part of the FOV). Moving towards upper layers, i.e. Ca~{\sc ii} H, this connection in the shape of loops is more present, although it is still faint. In the case of the O~{\sc i} lines, the magnetic loops that connect these concentrations are easier to detect. The pattern for Mg~{\sc ii} $k$ is similar to that of the O~{\sc i} lines although we can see the presence of more haze and less contrast \citep[see, for instance,][]{Leenaarts2013b}. Finally, the spatial distribution of the Lyman~$\beta$ line core intensity depicts the large scale loops that can be seen in Figure 11 of \cite{Carlsson2016}, demonstrating that the spectral line forms higher than the rest.

\begin{figure*}
\centering
\includegraphics[trim=0 0 0 0,width=18cm]{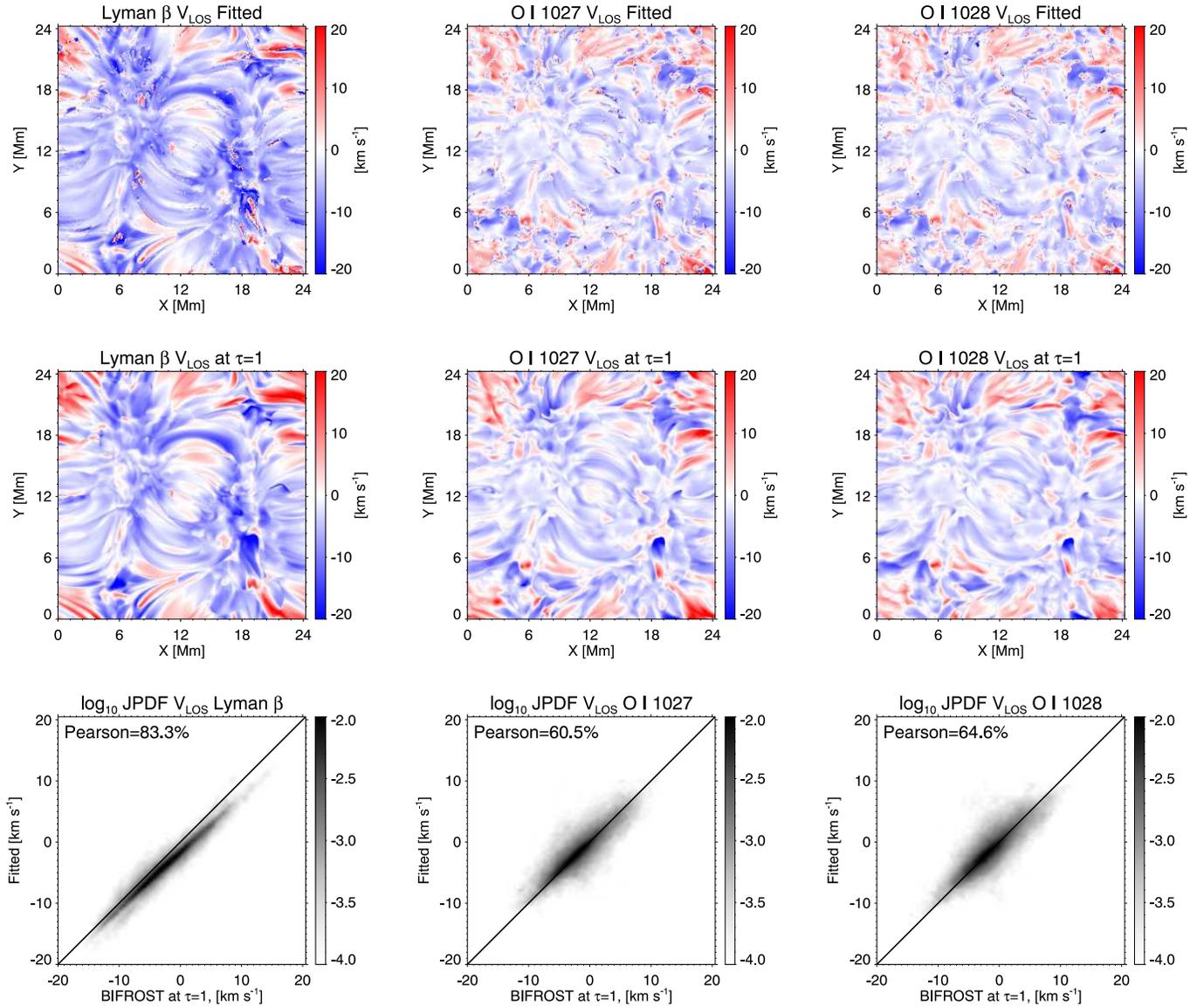}
\caption{Top row shows the LOS velocity inferred from the line core Doppler shift of, from left to right, ${\rm Lyman\beta}$, O~{\sc i} 1027, and 1028~\AA. Middle row displays the LOS velocity at the heights where the line core optical depth is unity. Bottom row shows the joint probability density function and the Pearson correlation between both LOS velocities. Solid black lines represent $y=x$. \label{fig:doppler}}
\end{figure*}

\subsubsection{Height of formation}

Figure~\ref{fig:heightbif} displays an estimation of the formation height for the spectral lines studied in the previous section. We represent for each pixel at $Y=12$~Mm (see, e.g., Fig~\ref{fig:bif_oi}) the height where the optical depth is unity for the line core wavelength. We also show the gas temperature in the background as a reference.

The formation height of Lyman~$\beta$ follows the transition region that in this cut approximately corresponds to the layer whose temperature is about $2\times 10^4$~K.  The heights where the optical depth is unity abruptly change along the selected cut (see pink) in agreement to what we showed in Figure~\ref{fig:bif_oi}.  We can find thin filaments that protrude up to 4~Mm and structures that can be as low as 2~Mm (often related to the presence of stronger magnetic fields). On average, the formation height of Lyman~$\beta$ in this cut of the simulation is about 2440~km.

Mg~{\sc ii}~$k$ forms close to Lyman~$\beta$ and always a bit lower than that line. In some cases, however, the differences are more significant. For instance, in the elongated structure at $X\sim 2.5$~Mm, Lyman~$\beta$ traces the upper edge of the thread while Mg~{\sc ii}~$k$ forms lower, around the central part of the protruding feature. This behavior indicates that both lines could complement each other if we aim to understand the properties of such structures.

Concerning the oxygen lines, they form lower in the atmosphere, with an average height of 1640~km and 1513~km for 1027 and 1028, respectively. In most cases, they form at similar heights, but at some locations, e.g. $X\sim12$ or $X\sim21$~Mm, the 1027 line reaches higher layers. We also included, as reference, an isothermal contour at 7000~K that seems to follow relatively closely the height where the optical depth is unity for these lines. In the case of the Ca~{\sc ii}, the near-UV transition forms at similar heights to those covered by the O~{\sc i} lines, while the infrared spectral line scans lower layers in the low-mid chromosphere. In the region $X=10\sim16$, magnetic loops are stratified and the transition region shifts to higher layers.  Following this trend, the spectral lines also form higher in the atmosphere, with the exception of Ca~{\sc ii} IR which is sensitive to the low chromosphere.

\subsubsection{Inferring the LOS velocity}

One of the main targets of future spectroscopic observations (e.g. done by Solar-C (EUVST) and SO/SPICE) is to examine the dynamics of the solar phenomena from the chromosphere to the outer atmosphere. We aim to estimate the capabilities of the spectral lines of interest to tackle this task. We do this by computing the Doppler shift of the line core of Lyman~$\beta$, and the two O~{\sc i} 1027 and 1028 \AA\ lines. 

We believe a single Gaussian fitting is not accurate enough for Lyman$\beta$ because the line shows multiple lobes, similar to the Mg~{\sc ii} UV spectral lines. Therefore, using the analysis of the IRIS observations as a reference, we perform the following steps. First, we detect the extrema of each spectral profile. Among them, we select the extremum which is nearest to the line core rest wavelength, and we perform a parabola fitting around the extremum to obtain the corresponding line core wavelength.

In the case of the O~{\sc i} transitions, we compute the line core Doppler shift using a single Gaussian fitting of the intensity profile.  We tried more complex procedures as we do for Lyman~$\beta$, but we did not achieve noticeable improvements.

The inferred LOS velocities for the entire snapshot FOV are plotted in the upper row of Figure~\ref{fig:doppler}. We include in the middle row the LOS velocity at the height where optical depth is unity for the line core wavelength (similar to what we did for the temperature in Figure~\ref{fig:bif_oi}). Additionally, we show in the bottom row the joint probability density functions (JPDF) of the velocity obtained by the fitting and that from the {\sc Bifrost} atmosphere (top and middle rows). In some pixels, the derived velocities were larger than 20~km/s. We believe those large values correspond to a bad fit of our method. The pixels where the velocity is larger than 20~km/s were excluded from the JPDF plots and from the computation of the Pearson correlation rate (although they are shown in the top row panels). Examining the JPDF results we can say that both methods provide a similar spatial distribution of the LOS velocity for the three spectral lines.  We can trace the velocities along the loop-like structures as well as inside the magnetic field elements. Differences among these LOS velocities can be seen in the JPDF plots and the Pearson correlation between both cases. For the hydrogen transition, the results are accurate with a correlation of 80\% while for the oxygen spectral lines, the correlation is lower being around 60\% for both transitions. It is worth mentioning that the fitting does not work well for some pixels where the LOS velocity shows opposite sign compared to that in the model atmosphere. 

\section{Discussion} \label{sec:discussion}

\subsection{Observability of the oxygen lines}

\begin{figure}
\centering
\includegraphics[trim=0 0 0 0,width=8.5cm]{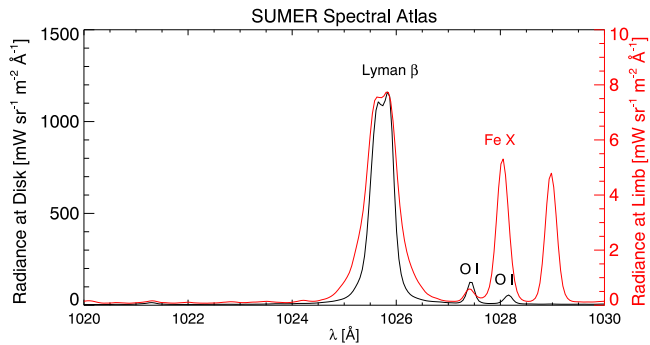}
\caption{SUMER spectral atlas for disk center (black) and limb (red) observations. \label{fig:atlas2}}
\end{figure}

In this work, we studied the capabilities of the oxygen lines to complement the Lyman~$\beta$ transition, improving the determination of the atmospheric parameters in the middle-upper chromosphere. We based our motivation on the fact that these lines are located in the wing of Lyman~$\beta$, clean of any other blended lines, as shown in Figure~\ref{fig:atlas}. However, when looking at the SUMER atlas at the limb, the situation is different (see red in Figure~\ref{fig:atlas2}). In this case, we can see the presence of coronal lines like Fe~{\sc x} that falls on top of 1028 impeding its observation. In other words, we could observe only 1027 when looking at low heliocentric angles. However, we do not see this as a significant limitation because we can always observe 1027 that is the one that forms highest in the atmosphere. Moreover, we could potentially improve our diagnostic capabilities including the coronal line because we would have in the same spectral window several lines that are sensitive to the atmospheric parameters at three distinct layers in the middle chromosphere, the transition region, and the corona.

\subsection{Improving the numerical set up}

We solved the radiative transfer equation in NLTE with hydrogen populations in statistical equilibrium. Departures from statistical equilibrium are important for the hydrogen ionization balance in a dynamic atmosphere \citep{Carlsson2002}. In future works, we plan on including these effects in a similar way as was done for He~{\sc i} in \cite{Golding2017} or for Lyman~$\alpha$ in \cite{Hong2019}.

\subsection{Diagnostic tools accuracy}

We developed simple techniques for inferring the LOS velocity from the Lyman and O~{\sc i} lines Doppler shifts. We tested them comparing the results with the velocities obtained at $\tau=1$ for the line core wavelength (i.e., the height where $\tau=1$ is highest). First, one could argue that this is not the best option as the spectral lines and the Doppler shifts that we derived from them do not form in a single atmospheric layer, and although we agree with that argument, we still consider that it is a fair comparison and one that is simple to understand. Thus, although we do not expect a one to one match, the results should be similar. In the case of the Lyman transition, the correlation is good, above $80\%$, so we can assume that our method is valid for this physical scenario and we could apply it on real observations from SoHO/SUMER or future missions. In the case of the O~{\sc i} transitions, the correlation is lower though. We indeed used a simplified method, directly assuming that the line is a single peak Gaussian, but we did not have indications to think otherwise. 
When checking the fits, we realized that the spectral profile, in most of the cases is a Voigt profile so we should continue assuming that a Gaussian fit is more than enough. However, we also found a strange shape on some profiles, and we are not sure about the cause yet. We plan to look further into this in the future checking whether they are related to a very complex atmosphere, to an inaccurate solution during the NLTE forward modeling or something else.
Finally, we did not include any spectral degradation in the process, which we believe sometimes leads to synthetic profiles with multiple  lobes at line core wavelengths. These multi-lobe intensity profiles are not present in the observations of, e.g., SoHO/SUMER. So, we assume that when including realistic conditions in terms of spectral and spatial resolution, most of those ``small-scale'' spectral variations will disappear. In future studies, we will concentrate on including instrumental smearing from missions like SoHO/SUMER, SO/SPICE, and Solar-C (EUVST), to confirm the previous assumption.

\section{Summary}

We examined the general properties of Lyman~$\beta$ and the O~{\sc i} 1027 and 1028~\AA \ lines located in its wing. We focused in the first part of this work on defining an appropriate synthesis configuration comparing the impact of NLTE and PRD on the lines of interest. We determined that the most balanced configuration between computational effort and accuracy is to simultaneously compute the H and O atomic populations in NLTE considering PRD effects for the Lyman~$\alpha$ and $\beta$ transitions. 

We studied the formation height and the sensitivity to the atmospheric parameters for the three spectral lines. We started with the semi-empirical atmosphere FALC, computing the height where the optical depth is unity for the entire spectral window at 1025~\AA. Lyman~$\beta$ forms at the transition region (the top boundary of the FALC atmosphere), while the 1027 and 1028 lines reach lower in the atmosphere at similar heights. The response function to temperature shows that Lyman~$\beta$ seamlessly covers from the middle chromosphere up to the transition region.  The 1027 and 1028 transitions are sensitive in a narrower range of heights from the middle to the upper chromosphere, with the 1027 line being sensitive to higher heights. In the case of the RF to LOS velocity, Lyman~$\beta$ is mainly sensitive to upper chromospheric and transition region layers through the wavelengths close to its line core. The O~{\sc i} lines, however, are sensitive at lower layers in the middle chromosphere. In all cases, we did not find any trace of degeneracy, i.e. a perturbation at a given height only affects a specific wavelength range.

We expanded these results examining one snapshot of the 3D enhanced network simulation computed with the {\sc Bifrost} code. Lyman~$\beta$ core intensity traces atmospheric loop-like features that connect the two opposite polarity magnetic field concentrations. The O~{\sc i} lines display a similar pattern but from lower atmospheric layers. 

We extended this study with two comparisons with traditional chromospheric lines like Ca~{\sc ii}~H and Mg~{\sc ii}~$k$. First, we examined the spatial distribution of line core intensity signals. Oxygen lines resemble the pattern showed by Mg~{\sc ii}~$k$ and seem to be sensitive to higher layers than those covered by Ca~{\sc ii}~H. Second, we computed the height where the optical depth is unity for the line core wavelength of these spectral lines. We used a slice of the simulation that crosses the magnetic loops that connect the two magnetic field concentrations (see $Y=12$~Mm in Fig~\ref{fig:bif_oi}). Lyman~$\beta$ traces the transition region, a corrugated layer that sometimes extends up to 4~Mm in the presence of cool plasma, followed closely by the Mg~{\sc ii}~$k$. The O~{\sc i} lines form lower in the atmosphere, at similar heights as those shown by the Ca~{\sc ii}~H. The infrared Ca~{\sc ii} spectral line is the one that forms deepest in the atmosphere, at 1~Mm on average.

Lastly, we estimated, in a simple fashion, the capabilities of the three spectral lines of interest for inferring the LOS velocities in the simulation. We determined that the Lyman transition is a good candidate for examining the LOS velocity along the loop-like structures that appear in the simulation and the oxygen spectral lines complement its capabilities scanning lower atmospheric layers. 
 
We plan to reinforce these conclusions analyzing observations from SoHO/SUMER studying different solar phenomena and pointing at different heliocentric angles. Also, we want to examine additional simulation snapshots where more complex chromospheric features are involved \citep[for instance,][]{Danilovic2017,Hansteen2017,Hansteen2019} to determine the capabilities of these lines to diagnose the energetics of such events. 

In summary, we have that the spectral window at 1025~\AA \ is an excellent candidate for understanding the thermal structure and plasma dynamics of the chromosphere and the transition region. Lyman~$\beta$ is sensitive to the temperature in a broad range of heights until the transition region. It can also cover the LOS velocities at these heights, although its sensitivity to lower layers is limited. However, the O~{\sc I} 1027 and 1028~\AA \ located in its wing fill the gap allowing a seamless height coverage from the middle chromosphere up to the transition region. In other words, future missions like SO/SPICE and Solar-C (EUVST) that are predominately focused on coronal activity will be able to trace the evolution of the coronal phenomena deep into the chromosphere and transition region when observing the spectral window at 1025~\AA. They will be able to, for instance, study the chromospheric and transition region dynamics on the onset of flares or the roots of solar wind.

\acknowledgments
We thank J. Leenaarts for providing the Mg~{\sc ii} atom that we used for different tests in this work. This work was supported by the Research Council of Norway through its Centres of Excellence scheme, project number 262622, and through grants of computing time from the Programme for Supercomputing, and by JSPS KAKENHI Grant Number JP18H05234 (PI: Y.
Katsukawa). This work is also supported by the Leading Graduate Course for Frontiers of 
Mathematical Sciences and Physics, The University of Tokyo, MEXT.

\bibliography{reference}{}



\end{document}